\documentclass[preprint,12pt]{aastex}

\newcommand{\Mfit}{M$_R^{fit}$}
\shorttitle{Imaging redshifts of BL Lac objects.}
\shortauthors{Sbarufatti et al.}
\begin{document}
\title{Imaging redshifts of BL Lac objects.}
\author{B. Sbarufatti\altaffilmark{1}, A. Treves}
\affil{Universit\`a dell'Insubria, Via Valleggio 11, I-22100 Como, Italy}
\author{R. Falomo}
\affil{INAF, Osservatorio Astronomico di Padova, Vicolo dell'Osservatorio 5,
 I-35122 Padova, Italy}
\altaffiltext{1}{also at Universit\`a di Milano-Bicocca}

\begin{abstract}

The HST snapshot imaging survey of 110 BL Lac objects \citep{urry00} has 
clearly shown that the host galaxies are massive and luminous ellipticals. 
The dispersion of the absolute magnitudes is sufficiently small, 
so that the measurement of the galaxy 
brightness becomes a valuable way of estimating their distance. This is
illustrated constructing the Hubble diagram of the 64 resolved objects with  
known redshift. By means of this relationship  we estimate the redshift of 
five resolved BL Lacs of the survey, which have no spectroscopic
$z$. 
The adopted method allows us also to evaluate lower limits to the redshift for 
13 objects of still unknown $z$, based on the lower limit on the host galaxy 
magnitude. This technique can be applied to any  BL Lac object
for which an estimate or a lower limit of the host galaxy  magnitude is available.
Finally we show that the distribution of the nuclear luminosity of 
all the BL Lacs of the survey, indicates that the objects for which both the 
redshift and the host galaxy are undetected, are among the most luminous, and 
possibly the most highly beamed.

\end{abstract}
\section{Introduction}

Contrary to the large majority of Active Galactic Nuclei (AGN), characterized
by optical spectra with prominent emission lines, BL Lac objects have quasi
featureless spectra.  In fact by the  definition of this class of AGN, the line
equivalent widths should be very small. The effect of this weakness of the
spectral features has the consequence to hamper in many cases the
determination of the redshift, thus making the distance of the source
difficult to assess. As originally proposed by \citet{blandford78}, the
weakness of the spectral lines is most probably due to the fact that the
underlying non thermal continuum is exalted by the relativistic beaming of a 
jet pointing in the observer direction.

There are two possible strategies for deriving the distance (redshift)
of BL Lacs. The first one is to improve the S/N ratio of the optical spectra, 
so that very weak spectral features may become detectable. 
At present some significant progress with respect to the existing data 
can be obtained through the use of 8m class telescopes 
\citep[e.g. ][]{heidt04, sbarufatti05a, sbarufatti05b, sowards05}.

The second approach requires high quality imaging of the objects, with the
scope of detecting the host galaxy. Using it as a standard candle one can
obtain an albeit indirect estimate of the distance 
\citep[e.g.][]{romanishin87,falomo96,falomo99,heidt99,nilsson03}. For this approach  one
needs a combination of superb angular resolution and high efficiency, in order
to detect and characterize the faint extended light of the galaxy from the 
bright nucleus. These conditions become mandatory at high redshifts, because of
the dimming of the surface brightness, which scales as (1+z)$^{4}$. 

The first imaging studies of BL Lac hosts were carried out using ground based
telescopes which produced a preliminary  characterization of the properties of 
host galaxies for various datasets \citep[e.g. ][]{abraham91,falomo96,wurtz96,
falomo99,heidt99,nilsson03}. These works have consistently shown that BL Lac 
hosts are
virtually all massive ellipticals  (average luminosity M$_R$ =--23.7 and
effective radius R$_e \sim$ 10 kpc, assuming H$_0$=50 km s$^{-1}$ Mpc$^{-1}$,
$\Omega$=0). A substantial progress in this area was achieved through
the use of HST images, and in particular by the BL Lac snapshot survey, 
carried out with WFPC2. This produced high quality homogeneous images for 110 
objects \citep{falomo97,urry99,scarpa00a,scarpa00b,urry00,falomo00,odowd05}.

In this paper we reconsider the results of the HST snapshot survey, with the
aim of fully exploiting the information relevant for the determination of
the distance of BL Lacs. The starting point is the analysis of the absolute
magnitude distribution of the hosts. We show that since this
distribution is relatively narrow one can indeed use the host luminosity as a
standard candle  to evaluate the redshift. This is used in particular to
determine the redshift for objects the host galaxy of which is detected, but
that still have  featureless spectra.  We then focus on the nuclei of the
sources, and show that the unresolved objects with pure featureless spectra
are likely the most luminous and possibly beamed objects of the class.

\section{HST Images of BL Lacs}\label{section:hst}

For this work we have considered the dataset of images obtained by the HST BL 
Lac survey discussed by \citet[][in the following addressed as U00]{urry00}. 
It contains 110 objects imaged with HST+WFPC2 in the F702W filter. 
The U00 sample was selectted from seven flux limited BL Lac
samples in literature (1Jy, PG, HEAO-A2, HEAO-A3, EMSS, Slew). It 
includes both high energy peaked (HBL) and low energy peaked (LBL) objects 
\citep[see][for HBL and LBL definition]{padovani95}, covering a wide range of 
jet physics in BL Lacs.
The surface brightness profile for each object was modeled with a combination 
of a point source (the nucleus component), and an extended emission (the host 
galaxy) represented by  an elliptical (de Vaucouleurs law) or a disk component
(exponential law). For all the sources where the host galaxy can be  detected
the radial  brightness profile is consistent with the host being an
elliptical. For the unresolved sources, U00 give a lower limit on the apparent 
magnitude of the underlying nebulosity. 

In order to construct a more homogeneous and updated dataset of BL Lac host
galaxies measurements we have introduced a number of revisions in the
treatment of the results of U00, which are listed in the following:
\begin{enumerate}
\item For three objects 0145+138, 0446+449, 0525+713 there is no evidence of a 
nuclear component in the HST images. This makes  the classification as BL Lac 
dubious, and therefore they are not considered here. The object 
1320+084 has been excluded from the sample, since recent spectroscopy
\citep{sbarufatti05b} has shown that the source is a broad line QSO 
at z=1.5. Our sample is then reduced to 106 objects;
\item For four objects new redshifts have been published in the last years:
0426-380, 1519-273 \citep{sbarufatti05a}, 0754+100; 
1914-194 \citep{carangelo03}, and for one ( 0158+001) the redshift was revised
(SDSS\footnote{see Sloan Digital Sky Survey Data Release 4, 
http://cas.sdss.org/astro/en/tools/getimg/spectra.asp, plate 
403/51871, fiber 631, and \citet{richards02} for a description of the
quasar survey});
\item The galactic extinction is now accounted for following
  \citet{schlegel98};
\item The K-correction for galaxy magnitudes is taken from \citet{poggianti97};
\item The adopted cosmological parameters are: H$_0$= 70 km s$^{-1}$ Mpc$^{-1}$, 
$\Omega_{\Lambda}$=0.7, $\Omega_m$=0.3;
\item In order to compare the luminosity of the hosts at different redshifts
 (and epochs) we have also included a correction for taking into account the
 passive evolution of the galaxies following \citet{bressan98}
 prescriptions. This  correction, weighted on the redshift distribution, is 
$\sim$0.3 magnitudes. 
\end{enumerate}
Table \ref{tab:urryrev} reports the relevant parameters for the 106
objects.
According to the spectra and imaging properties the objects in the 
sample can be divided in 4 groups:

\begin{itemize}
\item[A)]Objects for which the redshift is known from spectroscopy, and
  the host galaxy is detected (N= 64). All the sources with z$\leq$0.2  belong
  to this class \citep[see][]{falomo00};
\item[B)] Objects for which the host galaxy has been detected but with unknown 
redshift (N= 5);
\item[C)] Objects of known redshift but for which the host galaxy has not been 
detected (N=24);
\item[D)] Objects of unknown redshift and that have not been resolved by HST 
optical imaging (N= 13).
\end{itemize}

\begin{figure}[htbp]
  \epsscale{0.9} \plotone{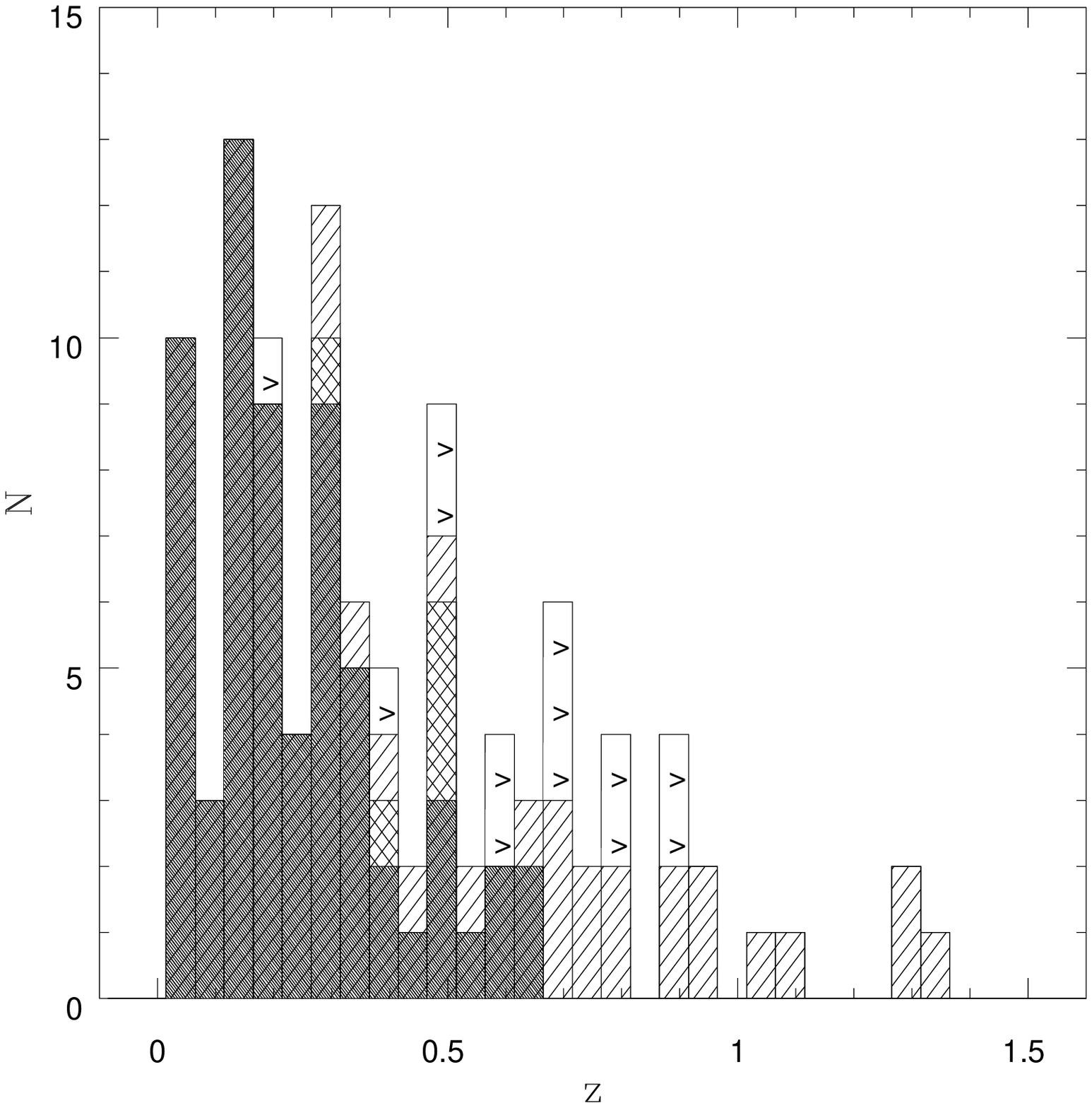}
  \caption{Redshift distribution of BL Lac objects in the HST snapshot imaging survey.
Group A (host detected, redshift known): shaded.
Group B (host detected, redshift unknown):  slashed.
Group C (host undetected, redshift known): open+ arrow. 
Group D (host undetected, redshift unknown): open. 
Lower limits are represented by arrows.}
  \label{fig:zdist}
\end{figure}

The redshift distribution of these groups is given in  Fig. \ref{fig:zdist}. 
As expected the objects in group A are clustered at low redshift (z $<$ 0.5) 
while those that are unresolved tend to be at z $>$ 0.5.

\section{Results}
\subsection{The Hubble diagram of the host galaxies of BL Lacs}

\begin{figure}[htbp]
  \epsscale{0.9} \plotone{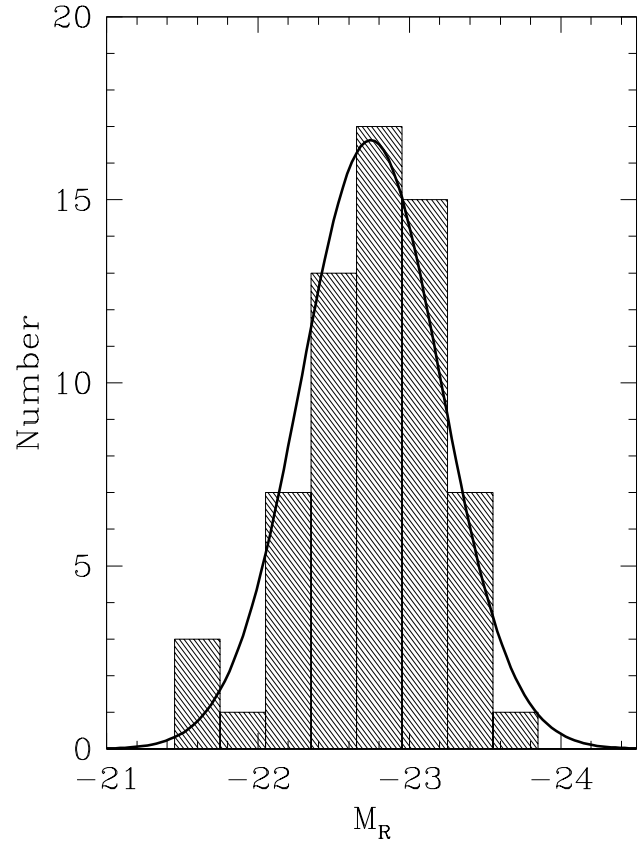}
  \caption{The distribution of the host galaxy absolute magnitude M$_R$\ for 
    objects of group A (host galaxy detected; redshift known). 
    The solid line represents a gaussian fit to the distribution (mean M$_R$=--22.8,
    $\sigma$=0.5).}
  \label{fig:distmh}
\end{figure}

The absolute magnitude of each host galaxy, modified for the 
effects of K and evolution correction \citep{poggianti97,bressan98} is 
reported in Table \ref{tab:urryrev}.
The distribution of the absolute magnitude (M$_R$)  of the host galaxies for 
objects in group A is reported in Fig. \ref{fig:distmh}. This distribution is
 rather narrow and can be well approximated by a Gaussian with mean value 
$<$M$_R>$=--22.8 and standard deviation $\sigma_M$=0.5. 
Note that U00 reported $<$M$_R>$=--23.7$\pm$0.6. The main difference is due to 
the difference in cosmological parameters  (U00 used H$_0$= 50 km s$^{-1}$ Mpc$^{-1}$, $\Omega$=0) 
and to the addition  of the correction for the passive evolution of the 
galaxies.  These variations also produce a smaller dispersion of the distribution 
(0.5 with respect to 0.6 given in U00).

In order to test how the host galaxy luminosity of BL Lacs can be used as 
a standard candle we have constructed the Hubble diagram. 
The relation between the apparent magnitude m$_R$ of the host galaxy and the
redshift $z$ is given by: 
\begin{equation}\label{eq:hubble}
\textrm{m}_R = \textrm{M}_R - \textrm{K}(z) + \textrm{E}(z) - 5 + 5 \textrm{log}(\textrm{d}(z)) 
\end{equation}

where M$_R$ is the host absolute magnitude, K($z$) is the K-correction, E($z$) is 
the evolution correction, d($z$) is the luminosity distance. With the 
assumptions described above for the cosmology and the K and passive 
evolution corrections, the only remaining free parameter is the host magnitude
 M$_R$. The best fit of the observed data yields \Mfit=--22.9. The Hubble diagram (m$_R$
vs. $z$) for the BL Lac hosts together with the best fit is shown in Fig. 
\ref{fig:hubble}. It is noticeable that $\sim$ 70\% of the points representing 
the resolved BL Lacs are encompassed within \Mfit$\pm$0.5 mag (i.e.  
--23.4$<$M$_R<$--22.4). This Hubble diagram can be used to obtain a photometric
redshift of the objects from the measurement of the host galaxy apparent 
magnitude, or a lower limit on $z$ if only a lower limit on the magnitude is 
available. In the redshift range considered here ($z\lesssim$0.7) the Hubble 
diagram can be well represented by the following expression:
\begin{equation}\label{eq:hubblefit}
\textrm{log}(1+z)=\ (0.293 * \textrm{m}_R^2 + 7.19 * \textrm{m}_R + 45.1)*10^{-2},
\end{equation} 
which approximates the Hubble diagram with a precision 
better than 1\%.

\begin{figure}[htbp]
  \epsscale{0.9} \plotone{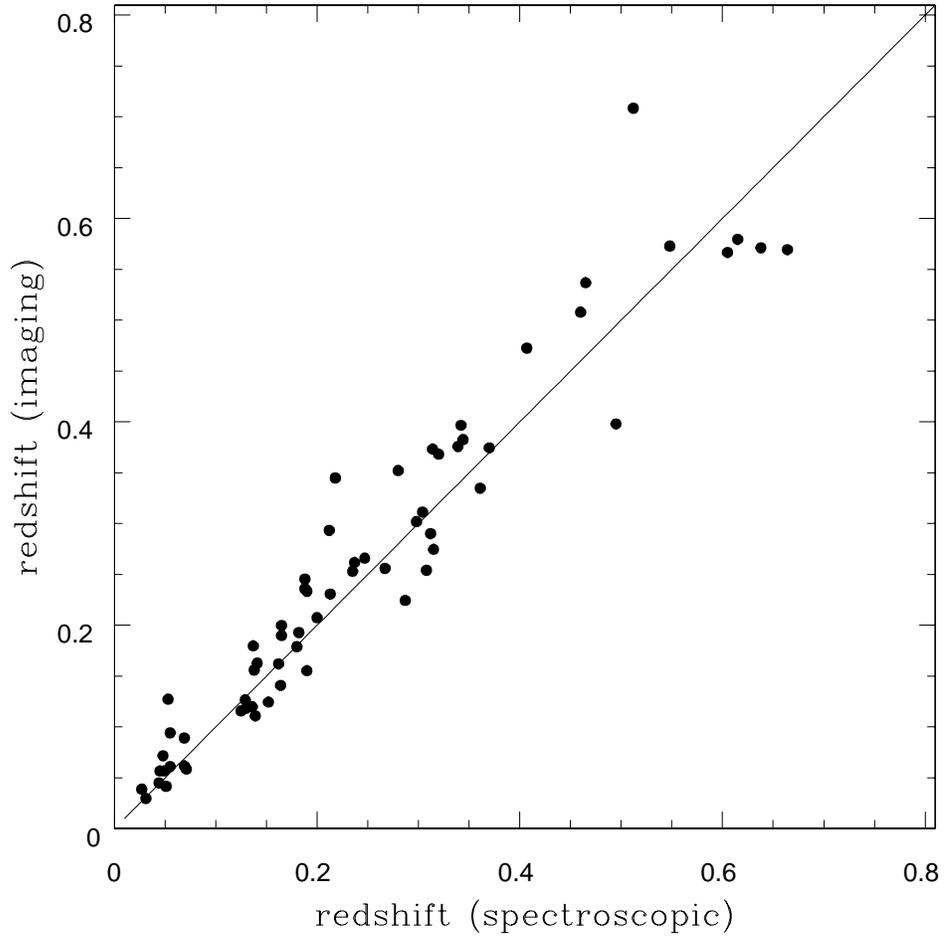}
  \caption{Comparison between imaging redshift obtained using the Hubble
    diagram fit and the spectroscopic redshift for the objects in group A (host
    detected, redshift known). The solid line represent the one to one relation.}
  \label{fig:zphotcomp}
\end{figure}

To evaluate the capability of this method we compare in
Fig. \ref{fig:zphotcomp} the redshifts given by this photometric technique with
the redshifts derived from the spectra in all cases where it is available.
The comparison shows that, apart very few exceptions, the $z$ estimated by the
host galaxy luminosity is in very good agreement with the one obtained
spectroscopically. The average difference of redshift between the two methods 
is: 
\begin{equation}\label{eq:fitrms}
<\Delta z> = 0.01 \pm 0.05 (rms).
\end{equation}

The main conclusion of this analysis is therefore that, at least in the
explored redshift range ($z <$ 0.7), the measurement of the apparent magnitude
of the host galaxy of a BL Lac source allows one to estimate its redshift with an 
average accuracy of 0.05. Because the U00 sample is unbiased as far as
host galaxies are concerned, this technique can be of use any time that the 
R apparent magnitude of a BL Lac host galaxy is measured, or a lower limit is obtained. 
In particular,
equation \ref{eq:hubblefit} gives a straightforward method to estimate the
redshift in such cases.
In the following sections we will use this 
technique to derive the redshift or a lower limit for the  
objects in groups B, C, and D.

\subsection{Imaging redshifts}

\paragraph{Group B (host galaxy detected, unknown spectroscopic redshift)}

Using the fit derived above for the Hubble diagram we can get a photometric
estimate of the redshift for the 5 objects with host detected and no
spectroscopic $z$ available (see Table \ref{tab:urryrev}; group B). 
The redshift of these objects ranges from $z$ = 0.26 to $z$ =0.54.
Combining the uncertainty deriving from equation (\ref{eq:fitrms}) with the 
one corresponding to the host galaxy apparent magnitude (which is of the order 
of 0.1 mag), the error on the imaging redshifts turns out to be 
$\lesssim$0.1. From the estimated $z$, one can then obtain the nucleus
luminosity. For the nucleus we adopted a K-correction following
\citet{wisotzki00}, under the hypothesis that its optical spectrum is described by 
a power law \citep[F$_{\lambda}\propto\lambda^{-\alpha}$\ with $\alpha$=0.7, see][]{falomo94}. 
Results are reported in Table \ref{tab:urryrev}.

\paragraph{Group C  (redshift known, host undetected)}

\begin{figure}[htbp]
  \epsscale{0.85} \plotone{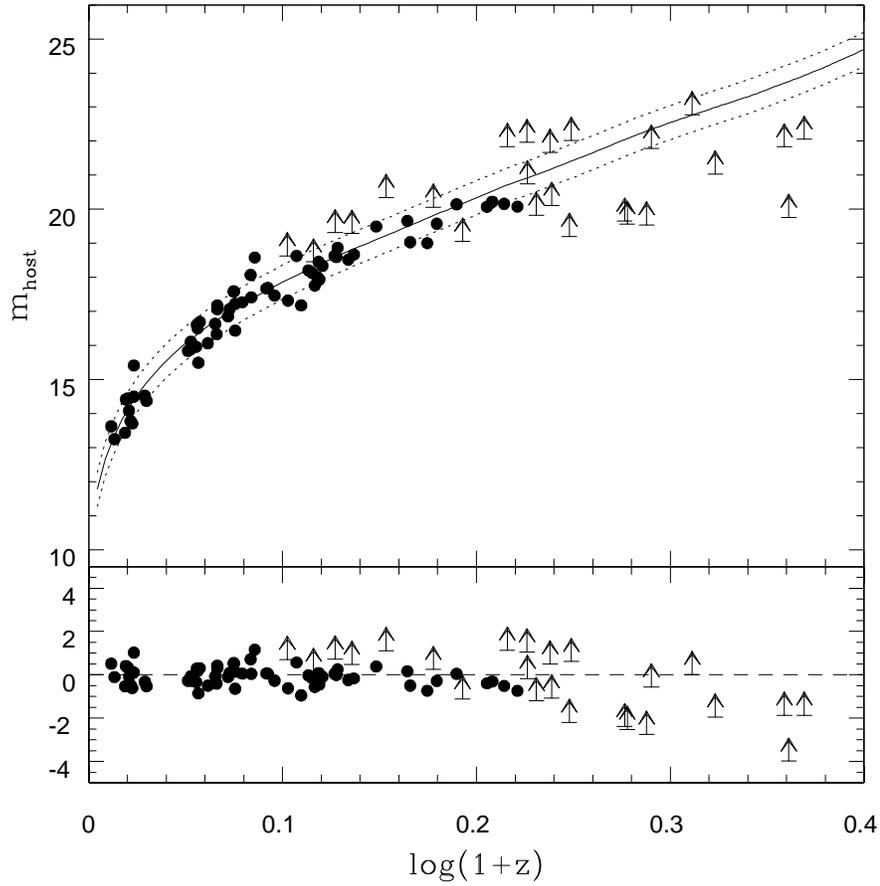}
  \caption{Hubble diagram for host galaxies of BL Lacs. Apparent magnitudes (R filter) are corrected
    for the galactic extinction.
    Group A (host detected, redshift
    known): filled circles. Group C (host undetected, redshift known) objects are
    shown as upper limits. The solid line corresponds to a fit with a galaxy of
    M$_R$=\Mfit=--22.9. Dotted curves correspond to a host galaxy 0.5 magnitudes
    brighter (lower curve) or fainter (upper curve). 
    Lower panel shows the deviations of the data from the fit.}
  \label{fig:hubble}
\end{figure}

\begin{figure}[htbp]
  \epsscale{0.85} \plotone{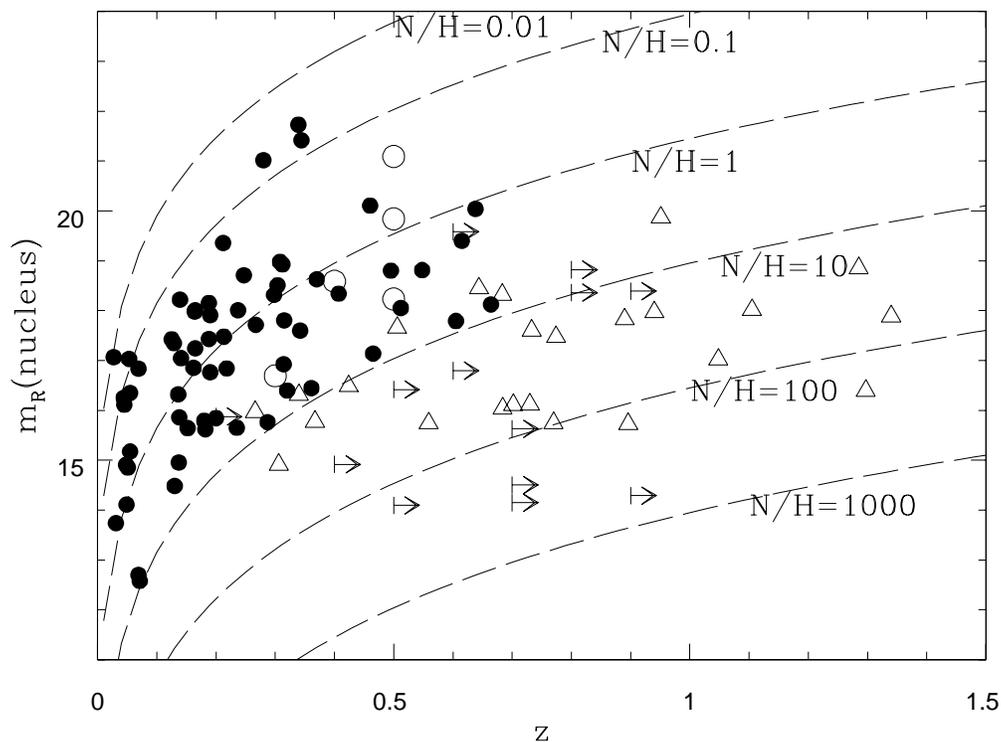}
  \caption{Distribution of nucleus apparent magnitude (corrected for 
    galactic extinction) vs. redshift. 
    Dashed lines show the loci of constant nucleus-to-host ratio (N/H), 
    assuming a host galaxy
    with M$_R$=\Mfit=--22.9.
    Group A (host detected, redshift known): filled circles.
    Group B (host detected, redshift unknown): open circles (redshift
    estimated using Hubble diagram.
    Group C (host undetected, redshift known): open triangles.
    Group D (host undetected, redshift unknown): arrows (lower limits on
    redshift determined using Hubble diagram).}
  \label{fig:mnnhz}
\end{figure}

From the HST images of these sources we have a lower limit on the magnitude of
the host reported by U00 (see Table \ref{tab:urryrev}) and we know the
redshift from the spectra. From these values we can obtain a lower limit on
the absolute magnitude of the host galaxy. These limits are in most cases
consistent (see Fig. \ref{fig:hubble}) with the presence of a host galaxy less
luminous than M$_R$=--23.4 (i.e. \Mfit--0.5). In no case the derived limit for the
host luminosity is lower than M$_R$=--21.9 (i.e. \Mfit+1.0). Therefore the fact that
the host galaxy has not been detected in these HST images is consistent with
the capability of the observation and with a distribution of the host absolute 
magnitude as given in Fig. \ref{fig:distmh}. The non
detection of the host in most cases is likely due to a high nuclear to host 
galaxy ratio (see section \ref{sec:discussion} and Fig. \ref{fig:mnnhz}).
        
\paragraph{Group D (no host, no redshift: extreme BL Lacs)}

The remaining group of objects is that formed by the unresolved sources which
exhibit a  featureless spectrum. The redshift of these objects is thus far
still unknown. The only information that can be derived from the images is
therefore the brightness of the nucleus and an upper limit to the brightness of
the surrounding nebulosity. Assuming that also these objects are hosted in a
galaxy with M$_R$=\Mfit=--22.9 and from the lower limit of the magnitude of
the surrounding nebulosity one can derive a lower limit to the redshift using
the Hubble diagram (see Fig. \ref{fig:hubble} and Table
\ref{tab:urryrev}). This lower limit to the redshift can then be used to
derive a lower limit to the luminosity of the nucleus. The distribution of the
absolute magnitude of the nuclei for all  objects in the four groups is shown
in Fig \ref{fig:distmn}. It turns out that most of the sources in group D
are more luminous than M$_R$=--25 and fill the bright tail of the 
luminosity distribution. It is worth to note also that of the 13 objects in 
this group, 10 are HBL and only 3 are LBL, while in the whole sample there are 
71 HBL and 35 LBL.

\section{Discussion:  nuclear luminosities and beaming.}\label{sec:discussion}

We have shown that the detection of the host galaxy allows a photometric
determination of the redshift for the 5 objects in group B. Since these
objects are at 0.2$<z<$0.5 they are convenient targets for redshift
measurement through spectroscopy.
Direct estimates of the absolute nuclear magnitude are available
for the objects with known spectroscopic redshift (group A and C). 
In addition, using the imaging redshifts we can add 5 more targets
(group B objects) and 13 upper limits for the objects in group D. The 
 redshift distribution of the whole sample is given in Fig. \ref{fig:zdist}. 
 
\begin{figure}[htbp]
  \epsscale{0.85} \plotone{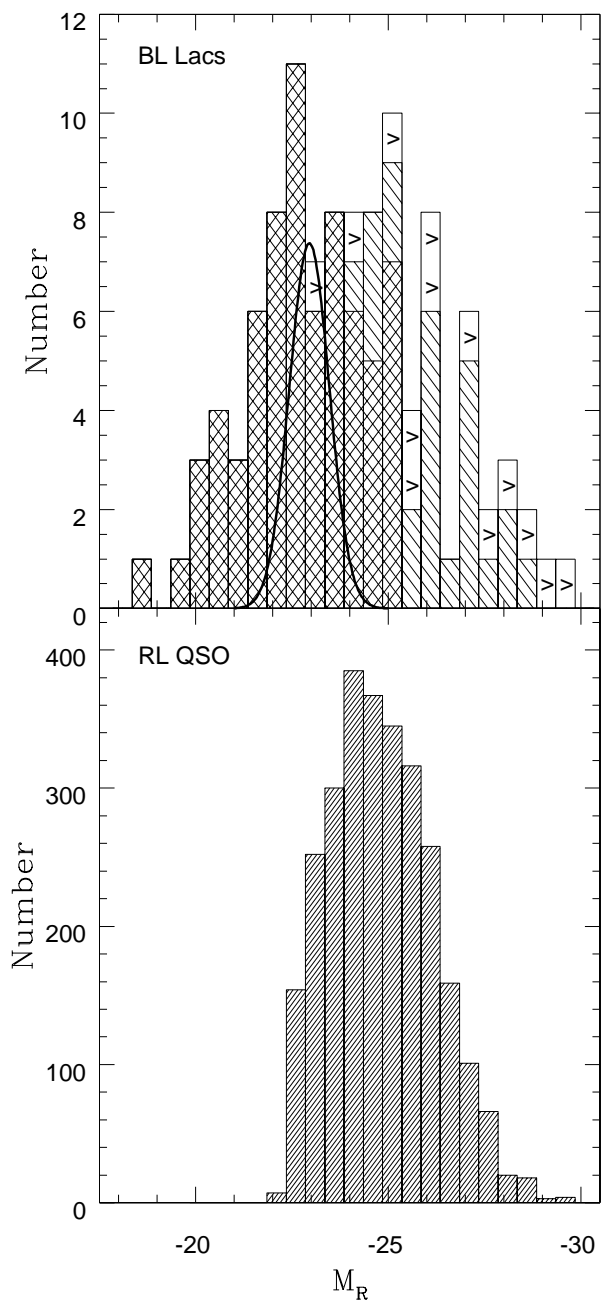}
  \caption{Upper panel (a):
    Distribution of nucleus absolute magnitude of BL Lacs in the R filter. 
    Group A+B (host detected): cross-hatched. 
     Group C (host undetected, redshift known): slashed. 
    Group D (host undetected, redshift unknown): open+arrow. 
    The solid line represents the gaussian fit to the host galaxy absolute 
    magnitude distribution for group A, scaled by a factor 0.5. 
    Upper limits are represented by arrows.
    Lower panel (b): Distribution of absolute R magnitudes for radio loud quasars (RL QSO) in \citet{veron03} catalogue with $z<$1.4.}
  \label{fig:distmn}
\end{figure}

The absolute nuclear magnitudes are summarized in Table \ref{tab:urryrev}. The 
mean and median absolute magnitude  for  objects in the groups A+B+C are  
M$_R$=--23.7 and M$_R$=--23.6 respectively, while adding objects in group D 
the median is M$_R$=--24.1. As already noted by U00 the distribution of the 
absolute magnitude of the host galaxies is much narrower than that of their 
nuclei (Fig. \ref{fig:distmn}).

Since the objects in each individual sample used to produce the HST snapshot 
survey are selected on the basis of their radio and X-ray fluxes, it is 
expected that the entire dataset of objects suffers from the typical bias of 
the flux limited surveys.
Nevertheless some interesting comments should be made. Our analysis of the
redshift and the nuclear luminosity of the objects clearly shows that the
{\it extreme} BL Lacs (objects with featureless spectra and unresolved; 
group D), are among the most luminous sources in the sample. Indeed 10 out of 
13 objects are brighter than the brightest object in the group A (see
Fig. \ref{fig:distmn}a). If one compares the luminosity distribution with that 
of z $<$ 1.4 radio loud quasars (using homogeneous corrections for the 
magnitudes) used for the BL Lacs  it is apparent that the bright tail of the 
BL Lacs distribution is consistent with that of normal radio loud quasars (see 
Fig \ref{fig:distmn}b).

Below the absolute magnitude M$_R$=--25 there are basically two types of
objects (groups C and D). Unless the latter are really significantly more
luminous than the average of objects in group C (objects with emission lines)
the different spectral properties could be related to different amount of
beaming. If one assumes that the two types of objects have a similar Broad Line
Region, the strength of the line should be related to the intrinsic ionizing
flux. If blazars behave as normal quasars the line emission should depend on
the unbeamed continuum \citep[e.g.][and references therein]{pian05}. Under
this hypothesis therefore we argue that in extreme BL Lacs the mechanism for
emission line formation is  less efficient because the intrinsic continuum
source is smaller with respect to group C objects, but more strongly beamed.

This suggestion is made stronger by the consideration of the distribution of 
the objects in the plane m$_{R}$\ vs. $z$  (see Fig. \ref{fig:mnnhz}).
The curves represent the loci of a constant Nucleus/Host ratio (N/H), defined 
as the ratio of monochromatic R band luminosities in the object rest frame. 
All the resolved sources are in the region with N/H less than 10 while the 
large majority of objects in groups C and D are in the region between N/H=10
 and N/H = 1000. Again in this context group D objects are clearly the most 
extreme, implying high N/H ratios (sometimes in excess of N/H=100). 
It is also noticeable that group D objects  are mostly of the HBL type , while
 the majority of group C objects are of the LBL type. The inference is 
therefore that in the optical band HBLs are more beamed than LBLs.

Independent information on the amount of beaming can be derived 
from the broad band spectral energy distribution, assuming that the emitted
continuum is basically the superposition of a synchrotron and an inverse 
Compton component \citep[see e.g.][]{ghisellini98}. Thus far these studies were
limited to objects of known redshifts, which is irrelevant to our proposal 
that extreme BL Lacs are extremely beamed objects, but some extension to
this class of sources may soon become available.

\section{Conclusions}

We have reanalyzed the host galaxies and nuclear properties of the BL Lac
objects observed by  HST snapshot imaging survey. The magnitudes of the
objects in the dataset have been revised according to the treatment of the
galactic extinction, the evolution and K-corrections. The concordant 
cosmological parameters have been used.

The main results and conclusions from this study are:

\begin{itemize}
\item The host galaxy absolute magnitude distribution is sufficiently narrow
  (gaussian distribution  centered around M$_R$=--22.9) that the host galaxy 
  can be used as a standard candle to derive the photometric 
  redshift of the objects. 
  Therefore a determination of $z$ (or a lower limit) can be simply derived 
  from the measurement (or from the lower limit) of the host galaxy apparent 
  magnitude.
      
\item The determination of the redshift and the lower limits allow us to 
  investigate the nuclear luminosity distributions for various type of objects. 
  This suggests that the objects in the sample that are unresolved and
  characterized by a featureless spectrum are  the most luminous  and /or
  beamed nuclei of the class. These extreme BL Lacs could have a 
  nucleus-to-host galaxy ratio of 100 to 1000.

\end{itemize}

\acknowledgments
This work was partially found by the Italian Space Agency (I/R/056/02) and the 
Italian Ministery of Education COFIN 2004023189.

\newpage

\newpage

\begin{deluxetable}{lcclrccllrrr}
\tabletypesize{\scriptsize}
\tablecaption{Host galaxy and nuclear properties of BL Lac
  objects\label{tab:urryrev}}
\tablehead{
\colhead{object}& 
\colhead{type}&
\colhead{group}&
\colhead{A$_R$}&
\colhead{z}&
\colhead{Galaxy}&
\colhead{Nucleus}&
\colhead{Evolution}&
\colhead{m$_n$}&
\colhead{M$_n$}&
\colhead{m$_h$}&
\colhead{M$_h$}\\
\colhead{}& \colhead{}& \colhead{}& \colhead{}& \colhead{}& \colhead{K corr.}&\colhead{K corr.}& \colhead{corr.}& \colhead{}& \colhead{}& \colhead{}& \colhead{}\\
\colhead{(1)}&
\colhead{(2)}&
\colhead{(3)}&
\colhead{(4)}&
\colhead{(5)}&
\colhead{(6)}&
\colhead{(7)}&
\colhead{(8)}&
\colhead{(9)}&
\colhead{(10)}&
\colhead{(11)}&
\colhead{(12)}\\}
\startdata
0033+595 &H    &D    &2.354  &[$>$0.24]& 0.71    & 0.06 &$-$0.39 &18.23& $<$$-$24.25   & $>$20.00 &$-$22.90\tablenotemark{a}\\
0118$-$272 &L    &C    &0.036  & 0.559   & 0.91    & 0.14 &$-$0.43 &15.78&    $-$26.96   & $>$19.09 &$>$$-$23.86 \\
0122+090 &H    &A    &0.249  & 0.339   & 0.43    & 0.10 &$-$0.28 &21.98&    $-$19.65   &  18.88   &$-$22.67 \\
0138$-$097 &L    &C    &0.079  & 0.733   &  1.4    & 0.18 &$-$0.61 &17.68&    $-$25.87   & $>$20.19 &$>$$-$23.84 \\
0158+001 &H    &A    &0.064  & 0.298   & 0.36    & 0.08 &$-$0.24 &18.38&    $-$22.72   &  18.27   &$-$22.75 \\
0229+200 &H    &A    &0.360  & 0.139   & 0.15    & 0.04 &$-$0.14 &18.58&    $-$20.93   &  15.85   &$-$23.54 \\
0235+164 &L    &C    &0.211  & 0.940   & 2.01    & 0.22 &$-$0.78 &18.18&    $-$26.21   & $>$19.75 &$>$$-$25.55 \\
0257+342 &H    &A    &0.468  & 0.247   & 0.29    & 0.07 &$-$0.21 &19.18&    $-$21.86   &  17.93   &$-$22.99 \\
0317+183 &H    &A    &0.369  & 0.190   & 0.21    & 0.06 &$-$0.17 &18.28&    $-$22.09   &  17.59   &$-$22.57 \\
0331$-$362 &H    &A    &0.050  & 0.308   & 0.38    & 0.09 &$-$0.25 &19.03&    $-$22.15   &  17.81   &$-$23.28 \\
0347$-$121 &H    &A    &0.125  & 0.188   & 0.21    & 0.06 &$-$0.17 &18.28&    $-$21.74   &  17.72   &$-$22.17 \\
0350$-$371 &H    &A    &0.020  & 0.165   & 0.18    & 0.05 &$-$0.16 &18.03&    $-$21.60   &  17.08   &$-$22.38 \\
0414+009 &H    &A    &0.314  & 0.287   & 0.35    & 0.08 &$-$0.24 &16.08&    $-$25.19   &  17.49   &$-$23.67 \\
0419+194 &H    &A    &1.476  & 0.512   & 0.79    & 0.13 &$-$0.40 &19.53&    $-$24.44   &  21.05   &$-$23.02 \\
0426$-$380 &L    &C    &0.066  & 1.105   & 2.44    & 0.24 &$-$0.94 &18.08&    $-$26.61   & $>$21.10 &$>$$-$24.80 \\
0454+844 &L    &C    &0.316  & 1.340   & 3.04    & 0.28 &$-$1.11 &18.20&    $-$27.30   & $>$22.37 &$>$$-$24.79 \\
0502+675 &H    &A    &0.406  & 0.314   & 0.39    & 0.09 &$-$0.26 &17.33&    $-$24.28   &  18.86   &$-$22.64 \\
0506$-$039 &H    &A    &0.220  & 0.304   & 0.37    & 0.09 &$-$0.25 &18.73&    $-$22.61   &  18.35   &$-$22.87 \\
0521$-$365 &L    &A    &0.105  & 0.055   & 0.06    & 0.02 &$-$0.06 &15.28&    $-$21.99   &  14.60   &$-$22.43 \\
0537$-$441 &L    &C    &0.101  & 0.896   & 1.89    & 0.21 &$-$0.74 &15.83&    $-$28.30   & $>$19.66 &$>$$-$25.32 \\
0548$-$322 &H    &A    &0.094  & 0.069   & 0.07    & 0.02 &$-$0.08 &16.93&    $-$20.68   &  14.62   &$-$22.89 \\
0607+710 &H    &A    &0.512  & 0.267   & 0.32    & 0.08 &$-$0.22 &18.23&    $-$23.06   &  17.83   &$-$23.35 \\
0622$-$525 &H    &B    &0.241  &[0.41]   & 0.54    & 0.11 &$-$0.33 &18.83&    $-$23.26   &  19.37   &$-$22.90\tablenotemark{a}\\
0647+250 &H    &D    &0.264  &[$>$0.47]& 0.48    & 0.11 &$-$0.33 &15.18& $<$$-$26.93   & $>$19.10 &$-$22.90\tablenotemark{a}\\
0706+591 &H    &A    &0.103  & 0.125   & 0.14    & 0.04 &$-$0.13 &17.53&    $-$21.54   &  15.94   &$-$22.95 \\
0715$-$259 &H    &A    &0.989  & 0.465   & 0.68    & 0.12 &$-$0.37 &18.13&    $-$25.08   &  20.02   &$-$23.23 \\
0716+714 &H    &D    &0.082  &[$>$0.52]& 0.71    & 0.13 &$-$0.39 &14.18& $<$$-$28.34   & $>$20.00 &$-$22.90\tablenotemark{a}\\
0735+178 &L    &C    &0.094  & 0.424   & 0.59    & 0.12 &$-$0.34 &16.58&    $-$25.50   & $>$20.44 &$>$$-$21.62 \\
0737+744 &H    &A    &0.073  & 0.315   & 0.39    & 0.09 &$-$0.26 &17.88&    $-$23.40   &  18.01   &$-$23.16 \\
0749+540 &L    &C    &0.111  & 0.730   & 1.39    & 0.18 &$-$0.61 &16.23&    $-$27.35   & $>$21.78 &$>$$-$22.26 \\
0754+100 &L    &C    &0.060  & 0.266   & 0.32    & 0.08 &$-$0.22 &16.03&    $-$24.80   & $>$18.69 &$>$$-$22.03 \\
0806+524 &H    &A    &0.118  & 0.138   & 0.15    & 0.04 &$-$0.14 &15.98&    $-$23.28   &  16.62   &$-$22.51 \\
0814+425 &L    &D    &0.170  &[$>$0.75]& 1.25    & 0.19 &$-$0.57 &18.99& $<$$-$24.90   & $>$21.47 &$-$22.90\tablenotemark{a}\\
0820+225 &L    &C    &0.112  & 0.951   & 2.04    & 0.22 &$-$0.79 &19.98&    $-$24.34   & $>$21.90 &$>$$-$23.36 \\
0823+033 &L    &C    &0.122  & 0.506   & 0.78    & 0.13 &$-$0.40 &17.78&    $-$24.79   & $>$20.18 &$>$$-$22.49 \\
0828+493 &L    &A    &0.117  & 0.548   & 0.88    & 0.14 &$-$0.43 &18.93&    $-$23.84   &  20.26   &$-$22.69 \\
0829+046 &L    &A    &0.087  & 0.180   &  0.2    & 0.05 &$-$0.17 &15.88&    $-$24.08   &  16.94   &$-$22.80 \\
0851+202 &L    &C    &0.076  & 0.306   & 0.38    & 0.09 &$-$0.25 &14.99&    $-$26.21   & $>$18.53 &$>$$-$22.57 \\
0922+749 &H    &A    &0.091  & 0.638   & 1.12    & 0.16 &$-$0.50 &20.13&    $-$23.04   &  20.25   &$-$23.25 \\
0927+500 &H    &A    &0.045  & 0.188   & 0.21    & 0.06 &$-$0.17 &17.48&    $-$22.46   &  17.62   &$-$22.19 \\
0954+658 &L    &C    &0.306  & 0.367   & 0.48    & 0.10 &$-$0.30 &16.08&    $-$25.81   & $>$19.60 &$>$$-$22.24 \\
0958+210 &H    &A    &0.061  & 0.344   & 0.44    & 0.10 &$-$0.28 &21.48&    $-$20.02   &  18.93   &$-$22.48 \\
1011+496 &H    &A    &0.033  & 0.200   & 0.23    & 0.06 &$-$0.18 &15.88&    $-$24.28   &  17.30   &$-$22.66 \\
1028+511 &H    &A    &0.033  & 0.361   & 0.47    & 0.10 &$-$0.29 &16.48&    $-$25.13   &  18.55   &$-$22.97 \\
1044+549 &H    &B    &0.038  &[0.54]   & 0.73    & 0.13 &$-$0.39 &19.88&    $-$22.59   &  20.05   &$-$22.90\tablenotemark{a}\\
1104+384 &H    &A    &0.041  & 0.031   & 0.03    & 0.01 &$-$0.03 &13.78&    $-$22.49   &  13.29   &$-$22.40 \\
1106+244 &H    &B    &0.045  &[0.46]   & 0.59    & 0.13 &$-$0.33 &18.28&    $-$24.20   &  19.57   &$-$22.90\tablenotemark{a}\\
1133+161 &H    &A    &0.173  & 0.460   & 0.67    & 0.12 &$-$0.37 &20.28&    $-$22.11   &  19.83   &$-$22.57 \\
1136+704 &H    &A    &0.035  & 0.045   & 0.05    & 0.01 &$-$0.05 &16.15&    $-$20.63   &  14.45   &$-$22.06 \\
1144$-$379 &L    &C    &0.257  & 1.048   &  2.3    & 0.23 &$-$0.88 &17.28&    $-$27.44   & $>$23.03 &$>$$-$22.82 \\
1147+245 &H    &D    &0.073  &[$>$0.63]& 0.95    & 0.15 &$-$0.46 &16.87& $<$$-$26.13   & $>$20.70 &$-$22.90\tablenotemark{a}\\
1207+394 &H    &A    &0.079  & 0.615   & 1.06    & 0.16 &$-$0.48 &19.48&    $-$23.59   &  20.30   &$-$23.05 \\
1212+078 &H    &A    &0.059  & 0.136   & 0.15    & 0.04 &$-$0.14 &16.38&    $-$22.82   &  16.02   &$-$23.02 \\
1215+303 &H    &A    &0.064  & 0.130   & 0.14    & 0.04 &$-$0.13 &14.55&    $-$24.64   &  15.99   &$-$22.95 \\
1218+304 &H    &A    &0.056  & 0.182   &  0.2    & 0.05 &$-$0.17 &15.68&    $-$24.25   &  17.12   &$-$22.62 \\
1221+245 &H    &A    &0.048  & 0.218   & 0.25    & 0.06 &$-$0.19 &16.89&    $-$23.41   &  18.63   &$-$21.56 \\
1229+643 &H    &A    &0.049  & 0.164   & 0.18    & 0.05 &$-$0.16 &18.03&    $-$21.63   &  16.38   &$-$23.09 \\
1239+069 &H    &D    &0.057  &[$>$0.92]& 1.63    & 0.21 &$-$0.75 &18.45& $<$$-$25.66   & $>$22.30 &$-$22.90\tablenotemark{a}\\
1246+586 &H    &D    &0.029  &[$>$0.73]& 1.14    & 0.17 &$-$0.57 &15.66& $<$$-$27.72   & $>$21.20 &$-$22.90\tablenotemark{a}\\
1248$-$296 &H    &A    &0.202  & 0.370   & 0.49    & 0.10 &$-$0.30 &18.83&    $-$23.02   &  18.87   &$-$22.89 \\
1249+174 &H    &C    &0.058  & 0.644   & 1.14    & 0.16 &$-$0.51 &18.51&    $-$24.66   & $>$21.90 &$>$$-$21.60 \\
1255+244 &H    &A    &0.034  & 0.141   & 0.15    & 0.04 &$-$0.14 &17.08&    $-$22.25   &  16.72   &$-$22.38 \\
1402+041 &H    &C    &0.069  & 0.340   & 0.43    & 0.10 &$-$0.28 &16.38&    $-$25.12   & $>$19.38 &$>$$-$22.00 \\
1407+595 &H    &A    &0.037  & 0.495   & 0.75    & 0.13 &$-$0.39 &18.84&    $-$23.59   &  19.04   &$-$23.47 \\
1418+546 &L    &A    &0.036  & 0.152   & 0.17    & 0.05 &$-$0.15 &15.68&    $-$23.82   &  16.10   &$-$23.18 \\
1422+580 &H    &C    &0.027  & 0.683   & 1.25    & 0.17 &$-$0.55 &18.35&    $-$24.95   & $>$21.99 &$>$$-$21.71 \\
1424+240 &H    &D    &0.156  &[$>$0.67]& 1.06    & 0.17 &$-$0.57 &14.66& $<$$-$28.85   & $>$21.00 &$-$22.90\tablenotemark{a}\\
1426+428 &H    &A    &0.033  & 0.129   & 0.14    & 0.04 &$-$0.13 &17.38&    $-$21.63   &  16.14   &$-$22.75 \\
1437+398 &L    &B    &0.037  &[0.26]   & 0.29    & 0.09 &$-$0.18 &16.73&    $-$24.43   &  17.95   &$-$22.90\tablenotemark{a}\\
1440+122 &H    &A    &0.076  & 0.162   & 0.18    & 0.05 &$-$0.15 &16.93&    $-$22.75   &  16.71   &$-$22.77 \\
1458+224 &H    &A    &0.128  & 0.235   & 0.27    & 0.07 &$-$0.20 &15.78&    $-$24.82   &  17.80   &$-$22.65 \\
1514$-$241 &H    &A    &0.369  & 0.049   & 0.05    & 0.02 &$-$0.05 &14.48&    $-$22.65   &  14.45   &$-$22.58 \\
1517+656 &H    &C    &0.068  & 0.702   &  1.3    & 0.17 &$-$0.58 &16.18&    $-$27.24   & $>$19.89 &$>$$-$23.94 \\
1519$-$273 &L    &C    &0.636  & 1.297   & 2.92    & 0.27 &$-$1.09 &17.03&    $-$28.69   & $>$20.40 &$>$$-$26.88 \\
1533+535 &H    &C    &0.051  & 0.890   & 1.87    & 0.21 &$-$0.74 &17.88&    $-$26.19   & $>$19.70 &$>$$-$25.19 \\
1534+014 &H    &A    &0.152  & 0.312   & 0.39    & 0.09 &$-$0.25 &19.08&    $-$22.27   &  18.16   &$-$23.08 \\
1538+149 &L    &A    &0.148  & 0.605   & 1.03    & 0.15 &$-$0.46 &17.94&    $-$25.14   &  20.22   &$-$23.14 \\
1544+820 &H    &D    &0.133  &[$>$0.46]& 0.60    & 0.13 &$-$0.33 &16.55& $<$$-$26.02   & $>$19.60 &$-$22.90\tablenotemark{a}\\
1553+113 &H    &D    &0.139  &[$>$0.78]& 1.31    & 0.21 &$-$0.68 &14.43& $<$$-$29.76   & $>$21.60 &$-$22.90\tablenotemark{a}\\
1704+604 &H    &A    &0.062  & 0.280   & 0.33    & 0.08 &$-$0.23 &21.08&    $-$19.92   &  18.69   &$-$22.15 \\
1722+119 &H    &D    &0.458  &[$>$0.68]& 1.23    & 0.17 &$-$0.57 &14.61& $<$$-$29.20   & $>$21.40 &$-$22.90\tablenotemark{a}\\
1728+502 &H    &A    &0.079  & 0.055   & 0.06    & 0.02 &$-$0.06 &16.43&    $-$20.81   &  15.49   &$-$21.51 \\
1738+476 &L    &D    &0.050  &[$>$0.60]& 0.87    & 0.15 &$-$0.39 &19.63& $<$$-$23.35   & $>$20.50 &$-$22.90\tablenotemark{a}\\
1745+504 &H    &B    &0.086  &[0.46]   & 0.59    & 0.13 &$-$0.33 &21.18&    $-$21.34   &  19.57   &$-$22.90\tablenotemark{a}\\
1749+096 &L    &A    &0.482  & 0.320   &  0.4    & 0.09 &$-$0.26 &16.88&    $-$24.88   &  18.82   &$-$22.81 \\
1749+701 &L    &C    &0.083  & 0.770   & 1.51    & 0.19 &$-$0.65 &15.83&    $-$27.87   & $>$19.28 &$>$$-$24.96 \\
1757+703 &H    &A    &0.089  & 0.407   & 0.56    & 0.11 &$-$0.33 &18.43&    $-$23.52   &  19.58   &$-$22.35 \\
1803+784 &L    &C    &0.140  & 0.684   & 1.25    & 0.17 &$-$0.56 &16.18&    $-$27.24   & $>$20.89 &$>$$-$22.91 \\
1807+698 &L    &A    &0.096  & 0.051   & 0.05    & 0.02 &$-$0.06 &14.95&    $-$22.30   &  13.87   &$-$22.97 \\
1823+568 &L    &A    &0.164  & 0.664   &  1.2    & 0.17 &$-$0.53 &18.29&    $-$25.07   &  20.24   &$-$23.49 \\
1853+671 &H    &A    &0.121  & 0.212   & 0.24    & 0.06 &$-$0.19 &19.48&    $-$20.88   &  18.19   &$-$21.99 \\
1914$-$194 &H    &A    &0.345  & 0.137   & 0.15    & 0.04 &$-$0.14 &15.30&    $-$24.19   &  16.95   &$-$22.39 \\
1959+650 &H    &A    &0.473  & 0.048   & 0.05    & 0.02 &$-$0.05 &15.38&    $-$21.85   &  14.92   &$-$22.17 \\
2005$-$489 &H    &A    &0.149  & 0.071   & 0.07    & 0.02 &$-$0.08 &12.73&    $-$25.23   &  14.52   &$-$23.11 \\
2007+777 &L    &A    &0.431  & 0.342   & 0.44    & 0.10 &$-$0.28 &18.03&    $-$23.84   &  19.03   &$-$22.73 \\
2037+521 &H    &A    &2.445  & 0.053   & 0.05    & 0.02 &$-$0.06 &19.48&    $-$20.12   &  16.15   &$-$23.12 \\
2131$-$021 &L    &C    &0.147  & 1.285   & 2.89    & 0.27 &$-$1.08 &19.00&    $-$26.20   & $>$21.98 &$>$$-$24.76 \\
2143+070 &H    &A    &0.200  & 0.237   & 0.27    & 0.07 &$-$0.20 &18.21&    $-$22.46   &  17.89   &$-$22.66 \\
2149+173 &L    &D    &0.270  &[$>$0.76]& 1.31    & 0.19 &$-$0.68 &18.63& $<$$-$25.36   & $>$21.60 &$-$22.90\tablenotemark{a}\\
2200+420 &L    &A    &0.880  & 0.069   & 0.07    & 0.02 &$-$0.08 &13.58&    $-$24.82   &  15.37   &$-$22.93 \\
2201+044 &L    &A    &0.113  & 0.027   & 0.03    & 0.01 &$-$0.03 &17.18&    $-$18.54   &  13.74   &$-$21.72 \\
2240$-$260 &L    &C    &0.057  & 0.774   & 1.53    & 0.19 &$-$0.65 &17.53&    $-$26.15   & $>$22.08 &$>$$-$22.17 \\
2254+074 &L    &A    &0.176  & 0.190   & 0.21    & 0.06 &$-$0.17 &16.94&    $-$23.24   &  16.61   &$-$23.36 \\
2326+174 &H    &A    &0.150  & 0.213   & 0.24    & 0.06 &$-$0.19 &17.63&    $-$22.76   &  17.56   &$-$22.66 \\
2344+514 &H    &A    &0.577  & 0.044   & 0.04    & 0.01 &$-$0.05 &16.83&    $-$20.49   &  14.01   &$-$22.98 \\
2356$-$309 &H    &A    &0.036  & 0.165   & 0.18    & 0.05 &$-$0.16 &17.28&    $-$22.37   &  17.21   &$-$22.27 \\
\enddata
\tablecomments{
(1) Object name;
(2) Object type. H: HBL, L: LBL;
(3) Object group. A: host galaxy detected, redshift known, B: host detected,
redshift unknown, C: host not detected, redshift known, D: host not detected,
redshift unknown;
(4) Galactic extinction coefficient by \citet{schlegel98};
(5) Redshift;
(6) Host galaxy K-correction, by \citet{poggianti97};
(7) Nucleus K-correction, calculated assuming F$_{\lambda}\propto\lambda^{-0.7}$;
(8) Host galaxy evolution correction;
(9) Nucleus apparent R magnitude;
(10) Nucleus absolute R magnitude;
(11) Host galaxy apparent R magnitude;
(12) Host galaxy absolute R magnitude.
}\tablenotetext{a}{Host galaxy absolute magnitude is assumed to be M$_R$=\Mfit=--22.9
  in order to obtain a redshift estimate}
\end{deluxetable}

\end{document}